# A Lateral MOS-Capacitor Enabled ITO Mach-Zehnder Modulator for Beam Steering

Rubab Amin, Rishi Maiti, Jonathan K. George, Xiaoxuan Ma, Zhizhen Ma, Hamed Dalir, Senior *Member, IEEE,* Mario Miscuglio, and Volker J. Sorger, Senior *Member, IEEE*

*Abstract*— Here, we experimentally demonstrate an Indium Tin Oxide (ITO) Mach-Zehnder interferometer heterogeneously integrated in silicon photonics. The phase shifter section is realized in a novel lateral MOS configuration, which, due to favorable electrostatic overlap, leads to efficient modulation ($V_\pi L = 63$ V.µm). This is achieved by (i) selecting a strong index changing material (ITO) and (ii) improving the field-overlap as verified by the electrostatic field lines. Furthermore, we show that this platform serves as a building block in an end-fire silicon photonics optical phased array (OPA) with a half-wavelength pitch within the waveguides with anticipated performance, including narrow main beam lobe (<3°) and >10 dB suppression of the side lobes, while electrostatically steering the emission profile up to ±80°, and if further engineered, can lead not only towards nanosecond-fast beam steering capabilities in LiDAR systems but also in holographic display, free-space optical communications, and optical switches.

## I. INTRODUCTION

Indium tin oxide (ITO) is a ternary compound which belongs to the class of transparent conductive oxide (TCO). For its coexisting optical and electrical properties, ITO has been extensively used as conductive layer in smartphones [1] and in photovoltaic cells [2]–[4]. When electrically tuned (i.e. capacitively gated), ITO films are able to deliver unity-strong index modulation [5]–[9], especially close to its epsilon-near-zero (ENZ) behavior [10], which leads, inter alia, to significant optical nonlinearities [11], supporting both strong index modulation [12], and slow-light effects [13].

The large interest surrounding ITO, which was mainly pushed due to its use in the industry, product-related purposes and timely applications, enabled high-yield and reliable wafer scale fabrication processes [14], [15], potentially compatible with the CMOS technology production line, which makes this material even more appealing for a plethora of other applications.

In this view, here we experimentally demonstrate a straightforwardly implementable ITO-based Mach Zehnder interferometer (MZI) electro-optic (EO) modulator utilizing a plasmonic mode. Thanks to favorable electrostatics in the active device region, the overlap of the in-plane component of the electrostatic field is maximized within the ITO layer, yielding an efficient carrier accumulation, thus a pronounced local refractive index variation. This mechanism of this initial prototype device enables an extinction ratio (ER) of 2.2 dB in a rather compact device (phase shifter length, L < 2 µm) and competitive figure of merit, $V_\pi L = 63$ V·µm, which suggest that these demonstrated device concepts trades-off rather well between the size (L) and the voltage needed to obtain a π-phase shift ($V_\pi$).

As a straightforward application, this active module could be particularly practical for targeting light detection and ranging (LiDAR) related device. LiDAR is a remote sensing method that uses laser pulses for high precision, long-range motion sensitive detection, which can resolve fine features, largely employed in self-driving cars [16], [17], and geographical [18], archeological [19], and geo-spatial surveys [20]. Although, to be able to scan a sufficiently large field of view a collection of antennas (i.e. free-space emitting waveguides), namely optical phased array (OPA), is used. Similar to their radio-frequency counterpart, by controlling the phase of the emission of each antenna a consistent signal is steered in a specific direction. This allows replacing the otherwise costly and fragile micro-electromechanical devices with integrated photonic systems [21], [22], working in the telecom regime (λ = 1550 nm), with reduced human eye exposure even at sustained power. Currently, the dense integration in silicon photonics enables the fabrication of a variety of OPAs for both mono-dimensional and bi-dimensional steering capabilities [23]. 1-dimensional beam steering can be achieved easily through end-fire sub-wavelength pitch array of waveguides; while 2-dimensional steering requires structure-aided emission in the direction orthogonal to the propagating wave, such as grating couplers [24], optical antennas [25], or heterogeneously integrated photonic crystals [26], [27]. However, the main critical aspect to these systems is that they rely on thermo-optical mechanisms for phase shifters which sets limitation in terms of steering rate; besides the implication to the power budget of the system, these devices are still operating at a speed in the kHz range [28], which cannot meet the requirement for robotic-based LiDAR

V. S. is supported by the Air Force Office of Scientific Research (AFOSR) (FA9550-17-1-0377) and Army Research Office (ARO) (W911NF-16-2-0194), and by AFOSR SBIR Phase II (FA9550-19-C-0003).

R. Amin, R. Maiti, J. K. George, X. Ma, Z. Ma, M. Miscuglio and V. J. Sorger are with Department of Electrical and Computer Engineering, The George Washington University, 800 22nd St. NW, Washington, DC 20052, USA (Corresponding author e-mail: sorger@gwu.edu).

H. Dalir is with Omega Optics, Inc., 8500 Shoal Creek Blvd., Bldg. 4, Suite 200, Austin, TX 78757, USA.

systems, aerospace applications and high-speed communication, while still potentially affected by thermal crosstalk.

Thanks to the fundamentally low RC delay, the modulation speed of ITO based film can achieve bandwidth up to GHz [5], [6]. Here, we use the experimental data of the ITO electro optic modulator to engineer a one-dimensional edge-emitting array of silicon waveguides and studying its performance. Using a $\lambda/2$ spacing (775-nm pitch) at the output, most of the power is conveyed through the main lobe, suppressing grating lobes [24], [29]. We show that the emission is characterized by a narrow profile in its main beam lobe (FWHM < 10°) and more than 10 dB suppression of the side lobes, while steering up to ±80°. We firmly believe that if this prototype phase shifter is further engineered it could be a promising approach towards a high-speed and energy-efficient beam steering platform for future LiDAR systems.

## II. Recent Advancements in Mach Zehnder Modulators

Since phase modulators invariably require interferometric schemes (e.g., ring resonators and MZIs), they do inherently suffer from an extended footprint compared to absorption modulators as those can be realized utilizing just straight waveguides. In MZMs, the product of the half-wave voltage times the active modulator length, $V_\pi L$, is a figure of merit (FOM) since they exhibit a tradeoff between obtaining π-phase shift with competing effects in increasing device lengths or bias voltages.

While MZMs based on lithium niobate (LiNbO₃) are commercially available, their $V_\pi L$ is rather high due to the weak Pockels effect, whereas improved performance is obtained with the quantum confined Stark effect (QCSE) in III-V semiconductors or emerging materials such as polymers and enhanced light-confinement for improving optical and RF mode overlap (Γ) with the active material (Table I). The devices exhibiting advanced FOMs amount to plasmonics, integration of organic/polymer materials, III-V quantum well structures, etc. Many of these schemes essentially offer acceptable performance but are mostly difficult to integrate in the mature Si process. Our design can avail ease of fabrication and potential CMOS integration due to the recent allowance of ITO in (certain) semiconductor foundry processes.

Among the LiNbO₃ modulators, three broad categories of schemes can be found: thin plate, ridge waveguides and domain inversion. Ridge waveguides provide step index contrast in the lateral direction. Thin plate LiNbO₃ modulators are formed by uniformly thinning down to dimensions of the EO crystal by precise lapping and polishing on a low dielectric substrate (e.g. SiO₂) to decrease the effective microwave index such that the modulating electric field is subjected to a higher confinement and forced to be parallel to the crystal z-axis exploiting the maximum EO tensor component ($r_{zz}$). Domain inversion employs opposite phase change in the two arms by unifying the waveguides under the same electrode diminishing chirp effects. While all these schemes do offer several necessary benefits, in terms of the key FOM they are quite limited (Table I). Different approaches using Si as the conventional material of choice have also emerged over the years such as simple metal-oxide-semiconductor (MOS) capacitive gating, employing pn/pin-configurations, selective doping optimizations etc. Different ambitious architectures (e.g. pipin [35], p₊-i-n₊ [52]) and techniques (e.g. doping optimization [39], projection structures [45], etc.) on Si based schemes have also been investigated in order to achieve higher modulation performances. The Si p₊-i-n₊ forward biased junction device was in the lead of minimizing the attainable FOM until 2011 [43]. III-V materials and quantum wells have also been sought after to bring down the FOM in realizing higher performing devices [46], [48], [50]. Hybrid structures using different material and structure combinations have also been investigated [49]. Recent use of organic polymers and hybridization with plasmonic structures enabled record-low FOMs [54], [55]. Previously we demonstrated an ITO-based photonic MZI and achieved a modest $V_\pi L$ despite it being the first of its kind [51]. This paper focuses on a lateral ITO-based MOS-stack enabled MZM device and the achieved FOM in our results seems relatively aligned with the state of the art MZMs and potentially paves the path for dense on-chip packaging for various applications including not only beam steering capabilities in LiDAR systems but also in holographic display, free-space optical communications, and optical switches.

TABLE I
Figure of merit (FOM) comparison for Mach Zehnder devices with different active modulation materials and waveguide structures in recent years

| Structure/Material | $V_\pi L$ (V.μm) | Ref. |
|---|---|---|
| Si Wrapped around-pn | 140,000 | [30] |
| Coplanar waveguide LiNbO₃ | 120,000 | [31] |
| Si Wrapped around-pn | 110,000 | [30] |
| Domain inverted push-pull LiNbO₃ | 90,000 | [32] |
| Dual driven coplanar waveguide LiNbO₃ | 80,000 | [31] |
| Si Vertical-pn | 40,000 | [33] |
| Bulk LiNbO₃ physical limit | 36,000 | [34] |
| Si pipin | 35,000 | [35] |
| Si Lateral-pn | 28,000 | [36] |
| Si Lateral-pn | 27,000 | [37] |
| Si pn-depletion | 24,000 | [38] |
| Doping optimized Si | 20,500 | [39] |
| Si Self-aligned-pn | 18,600 | [40] |
| Integrated thin film LiNbO₃ on insulator | 18,000 | [41] |
| Si pin | 13,000 | [42] |
| Silicon-organic hybrid (SOH) | 9,000 | [43] |
| Si Lateral-pn | 8,500 | [44] |
| Si Projection MOS | 5,000 | [45] |
| III-V Multiple Quantum Wells (MQW) | 4,600 | [46] |
| SOH | 3,800 | [47] |
| GaAs/AlGaAs | 2,100 | [48] |
| Hybrid Si MQW | 2,000 | [49] |
| InGaAlAs/InAlAs MQW | 600 | [50] |
| ITO MOS | 520 | [51] |
| Si p₊-i-n₊ | 360 | [52] |
| EO Polymer Plasmonic | 70 | [53] |
| ITO Lateral MOS | 63 | This work |
| Liquid crystals with SOH slot/ all-plasmonic polymer | 60 | [54], [55] |

## III. ITO Material Properties

The ITO material of our modulator parameters were derived by a combination of electrical and ellipsometric spectroscopic (J. A. Woollam M-2000 DI) measurements of the film, as deposited. Dispersion relations for the real and imaginary parts of the complex refractive index, $n$ and $\kappa$, respectively, with regards to states of operation of the device are shown in Fig. 1. The dispersion relations are analytically modeled with fitting parameters obtained from experimental ellipsometry. Hereafter, the nomenclature regarding the ON-OFF states of operation for the modulator relates to the light transmission (ON) vs. low-transmission (OFF) characteristics through the device, rather than the applied voltage bias. A 10 nm thin film is deposited on a $SiO_2$ substrate using the same fabrication processes as the actual device (details in sec. IV), contact pads of the same dimensions as present in the device were written using electron beam lithography (EBL) and subsequent Ti/Au deposition, and liftoff was also performed similar to the actual device. ITO resistivity and sheet resistance was found using the 4-point probe system to be approximately 250 $\Omega$ and 74 $\Omega/\square$, respectively. The resistivity and mobility of the ITO film was derived by resistor-transmission line measurements and confirmed by ellipsometry, therefore measured to be $6.2 \times 10^{-4}$ $\Omega$.cm and 37.6 cm2/V-s, respectively. Hall bar tests revealed that the carrier concentration of the as deposited ITO film is $N_c = 2.3 \times 10_{20}$ cm-3. The change in the carrier concentration level arising from active capacitive lateral gating is calculated as $\Delta N_c = 1 \times 10_{20}$ cm-3. Unlike other work on ITO absorption modulators [5], [6], this low carrier change was selected to reduce optical losses facilitating phase contributions in the obtainable modulation dynamic range.

## IV. Device Design, Fabrication and Results

We opted for a symmetrical passive MZI structure on a Silicon on insulator (SOI) platform so that the interference pattern at the output can distinguishably confer modulation effects from our active device. A symmetrical MZI structure here means that we choose the same length for both Mach-Zehnder arms and employ equal 50:50 Y-splitters on both sides. This design choice is made as different ratio in the Y-splitters can introduce chirp effects in modulation. Fabrication imperfections such as sidewall roughness, surface defects and alignment issues inherently deter one from achieving perfectly symmetrical MZI structures. Nevertheless, we make this design choice to minimize such effects and curtail the optical path length difference in both the arms of the MZI, to distinguish phase variability only conferring active modulation effects, while further engineering can be performed in an industry setting. The adverse effects of using highly-index tuning materials like ITO is accompanied by increased absorption as a byproduct of modulation relative to field-based modulation such as Pockels effect for example. This absorption in free-carrier materials (Si, ITO) arises from the well-known Kramers-Kronig (K-K) relations and poses a tradeoff for utilizing such highly tunable active modulation materials [56], [57]. As such, the active ITO in one arm of the MZI imposes a modulating absorption (loss) component, whereas the other (un-modulated) arm throughputs a higher optical power as a result. This, in turn, produces an arm loss imbalance in the device, thus limiting the achievable modulation dynamic range as differential loss between the arms due to carrier depletion can limit the ER [58].

For an ideal MZI the basic components i.e. the Y-splitters, arm lengths, arm losses and the phase shifters are perfect, and no issues with residual extinction ratio and chirp was found. In praxis, the Y-splitters do not have perfect equal power division and the phase shifts in both arms simultaneously do not conjugate each other. Let us model an MZI structure where the input Y-splitter power splitting ratio is $p_1^2 : (1 - p_1^2)$ and the output Y-junction power splitting ratio is $p_2^2 : (1 - p_2^2)$. Let us assume that in the two arms the corresponding phase shifts (regardless of whether due to active modulation or fabrication imperfections leading to mismatched arm lengths) are $\Delta\phi_1$ and $\Delta\phi_2$, respectively. If we now assume that the two arms of the interferometer lead to different propagation field losses $a_1$ and $a_2$ respectively, the output field, in general, can be expressed as

$$\tilde{E} = \widetilde{E_0}\left(a_1 p_1 p_2 e^{j\Delta\phi_1(t)} + a_2\sqrt{1-p_1^2}\sqrt{1-p_2^2}\, e^{j\Delta\phi_2(t)}\right) \quad (1)$$

The input and output fields are denoted by phasor quantities, i.e., $\tilde{E} = E e^{j\omega t}$, and assuming there is no gain in the system, $(a_1 + a_2)^2 \leq 1$. It is difficult to comprehend the effect of all the parameters on the extinction ratio and chirp from this expression. Let us, therefore, simplify this expression with the assumption that only the input Y-splitter is imperfect, every other component is ideal and the arm lengths and arm losses are balanced, which translates to $p_2^2 = 1/2$, $a_1 = a_2 = a$ and $|\Delta\phi_1 - \Delta\phi_2| = \Delta\phi$. With these approximations in place, we can obtain

$$\tilde{E} = \frac{\widetilde{E_0}}{\sqrt{2}} a e^{j\Delta\phi(t)}\left(p_1 + \sqrt{1-p_1^2}\, e^{j2\Delta\phi(t)}\right) \quad (2)$$

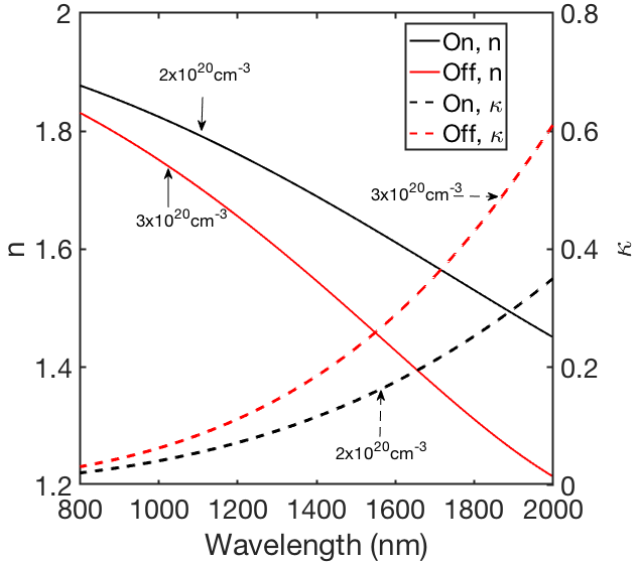

Fig. 1. ITO complex refractive index (real, $n$ and imaginary, $\kappa$ parts) as function of the wavelength in OFF ($3 \times 10_{20}$ cm-3, high carrier concentration) and ON state ($2 \times 10_{20}$ cm-3, initial carrier concentration). The dispersion relations are analytically modeled with fitting parameters obtained from experimental ellipsometry.

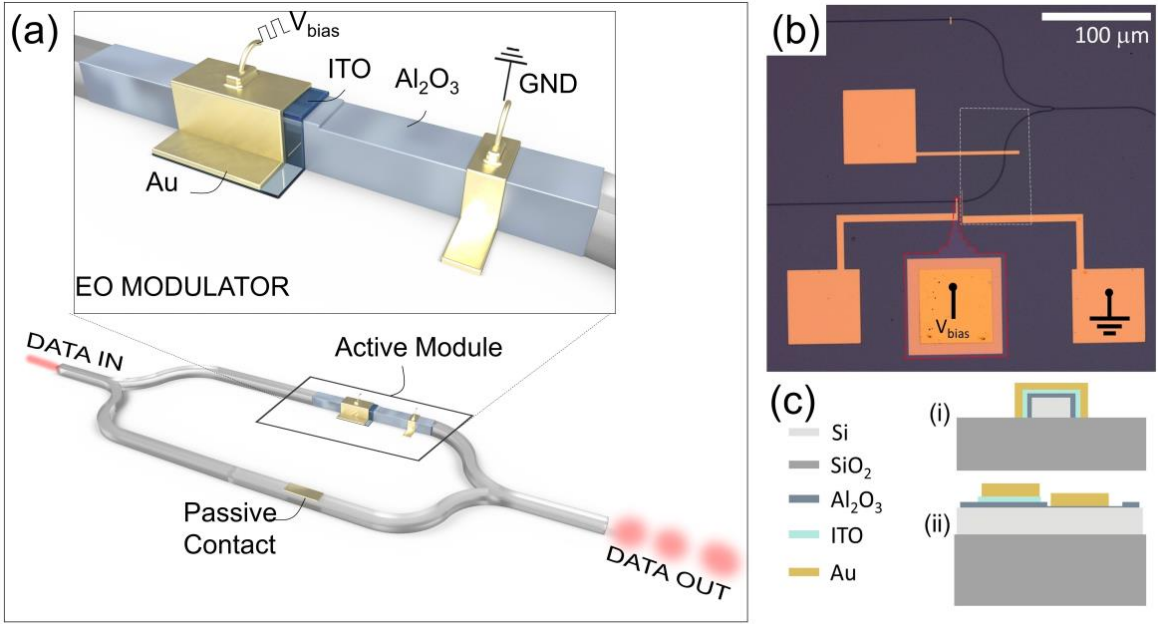

Fig. 2. (a) Schematic representation of the lateral MOS ITO based Mach Zehnder inteferometer operating at λ = 1550 nm; the phase shifter length (between the contacts) is <2 μm (b) Optical microscope image of the fabricated Mach Zehnder modulator with contact pads for biasing shown; the patterned ITO and the partial oxide etched region is highlighted with red and white dashes, respectively. (c) Schematic diagrams of the device in: (i) Cross-sectional view in the active ITO deposited region; and (ii) A longitudinal cross-section along the Si waveguide exhibiting the device region with the partial etched gate oxide region. The Al₂O₃ gate oxide is 10 nm thick, and partially etched down to 1-2 nm thickness in white dashed area in (b). Si waveguides are 500 nm × 220 nm, deposited ITO layer is 10 nm thick. Image not drawn to scale.

Now, the extinction ratio (ER) is the ratio of the transmission between the ON state (maximum transmission) and OFF state (minimum transmission), i.e., static ER since it is measured by varying a DC phase bias to one of the arms to find the absolute maximum and minimum transmission. This is necessary since the dynamic ER may be reduced when operating at high frequencies due to limited phase swings or pulse shaping from the finite bandwidth of the electrodes. This upper bound can be referred to the maximum extinction ratio, ER$_{max}$. For binary modulation schemes $\Delta\phi = \pm\pi/2$, and consequentially the extinction ratio becomes

$$ER = \frac{p_1 + \sqrt{1 - p_1^2}}{p_1 - \sqrt{1 - p_1^2}} \quad (3)$$

It is noteworthy that for an ideal MZI case, where $p_1^2 = 1/2$, the ER is infinite given that both arm losses are perfectly balanced notwithstanding the K-K resultant imbalance or fabrication imperfection leading to the different branches experiencing different optical path travelling in distinct waveguides. To investigate the effects of active modulation at the output, let the Y-junctions be ideal, i.e. $p_1^2 = p_2^2 = 1/2$ and all the field losses in the two arms are equal ($a_1 = a_2$). The output then becomes

$$\tilde{E} = a\widetilde{E_0}e^{j\Delta\phi(t)} = a\frac{\widetilde{E_0}}{2}e^{j\Delta\phi(t)}\cos^2\left(\frac{\Delta\phi}{2}\right) \quad (4)$$

To maximize the obtainable ER, i.e., ensuring minimal zeros in the OFF state, the field losses in both arms need to be matched, i.e., $a_1 = a_2$. Deviations from this ideal case are typically attributed to imperfect 50:50 Y-couplers [59], [60]. However, it is critical to emphasize that deviation from $a_1 = a_2$ can be a direct result from differences in the losses anywhere in the MZM configuration including possible fabrication imperfections. By contrast, higher index changeable materials

(e.g., ITO) do accompany loss as a byproduct of modulation and as such both the states of operation need to be accounted for in design considerations. One can improve this arm loss imbalance by tuning the un-modulated arm losses statically to counteract the imbalances arising from the K-K relations. As such, we chose to deposit metal (Au) on the other (un-modulated) arm of the MZ (Fig. 2(a)).

Since the modulation efficiency (ER/peak-to-peak voltage, V$_{pp}$) is improved for better electrostatics, we use a relatively high-k dielectric, a 10 nm oxide layer of Al₂O₃ grown on the passive structure via atomic layer deposition (ALD) for capacitive gating. Subsequently, a 10 nm thin film of ITO is deposited using an ion beam deposition (IBD) process after necessary patterning using EBL and liftoff processes afterwards (Fig. 2(b), red dashed area). The IBD process has synergies for processing ITO as this process yields dense crystalline films that are pinhole-free and highly uniform and allows for a room temperature process, which does not anneal ITO (i.e., no activation of Sn carriers as to facilitate electrostatic EO tuning). Incidentally, IBD technologies are advantageous for nanophotonic device fabrication due to their precise controllability of material properties such as microstructure, non-stoichiometry, morphology, and crystallinity [61], [62].

A selective etch step of the ALD grown oxide near the active ITO device region is enacted to facilitate the electric field overlap from the contacts with the active ITO material (Fig. 2(b), white dashed area). Contacts and the plasmonic top layer are formed by depositing 50 nm of Au using electron beam evaporation process. An adhesion layer of 3 nm of Ti is used in the process. The other contact is placed in close proximity (<2 μm) to the plasmonic top contact in the partial etched region to maximize the electrostatic field overlap to the active ITO region (Fig. 2(c, i)). The schematic of a longitudinal cross-section along the Si waveguide (active arm of the MZI structure) in the device region is illustrated in Fig. 2(c, i) and a cross-sectional schematic of the active plasmonic ITO region is shown in Fig. 2(c, ii). Another contact on the partial etched region is placed for determining the partial etch success; as we aimed for a remainder of just 1-2 nm thin oxide film after etch (Fig. 2(b)), this contact provided the control to determine if etched all the way through to the conductive Si layer.

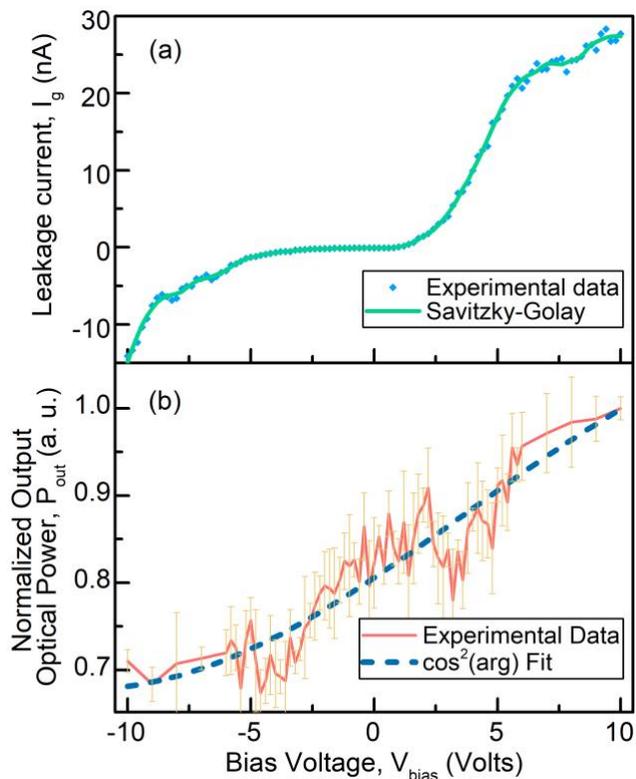

Fig. 3. (a) Normalized output optical power, $P_{out}$ (a. u.) vs. bias voltage, $V_{bias}$ (Volts) demonstrating the modulator performance, the experimental data was fit with a squared sinusoidal ($\cos_2(\text{arg})$) fit to extract the half-wave voltage, $V_\pi \approx$ 33 V. (d) I-V measurements of the ITO-oxide-metal lateral capacitive stack; a Savitzky-Golay smoothing function was applied on the experimental data to showcase the I-V characteristics.

The pattern transfers were performed in EBL using the Raith VOYAGER tool with PMMA based photoresists, and MIBK:IPA (1:3) developer for 60 s. 50 nm of Au for contacts and the plasmonic top layer in the mode structure were deposited using an e-beam evaporation system (CHA Criterion) as Au has reasonably low ohmic loss at near IR wavelengths. An additional 3 nm adhesion layer of Ti was used in the contacts. The $Al_2O_3$ oxide was deposited using the ALD technique as it provides reliable and repeatable performance characteristics. The Fiji G2 ALD tool was used at low temperature settings (100º C) for 100 cycles to deposit about 10 nm of $Al_2O_3$ to ensure higher film quality devoid of any pinholes or surface traps. A Filmetrics F20-UV system was used to characterize the $Al_2O_3$ deposition rate.

An etch step was required for the partial etch near the active device region to facilitate the necessary height contrast between the plasmonic top contact and the lateral bottom contact in the MOS-stack. We used a rather slow wet etch process for $Al_2O_3$ using an MF319 solution in the area of interest (near the active

ITO region, keeping the both contacts sufficiently close in proximity without jeopardizing etching on the extended active ITO region). Note, MF319 contains tetramethylammonium hydroxide (TMAH), which reacts with the Al and can etch the oxide thereof.

Experimental I-V measurements of the device show a working capacitor in the measured voltage range, not showing any observable saturation of the MOS capacitor or breakdown of the gate oxide characteristics (Fig. 3(a)). Electro-optic transmission power tests via a gated-transmission measurement exhibit reasonable modulation of the laser power demonstrating a modulation depth (i.e. ER) of ~1.34 dB in the measured bias range, and a squared cosine fit (as dictated by the underlying physics of MZIs from Eq. (4)) can obtain an ER of 2.2 dB. The quality of the fit symbolized by the coefficient of determination ($R_2$) is 0.86. The voltage needed for π-phase shifts at the optical output is about 33 V (Fig. 3(b)) and a corresponding $V_\pi L$ of just 63 V·μm given the <2 μm-short phase shifter.

In order to gain more insights into the field-distribution of this lateral-gated modulator, finite element method (FEM) simulations are carried out to resolve the electrostatic field overlap with the active ITO arising from capacitive gating which confirms an increased field overlap due to the partial

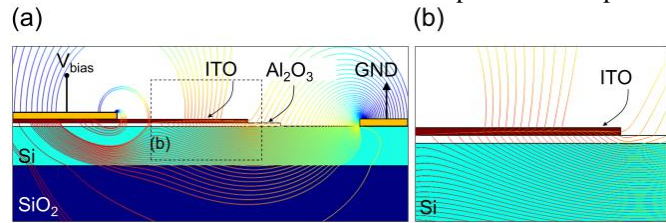

Fig. 4. FEM simulation of the longitudinal cross-sectional device region revealing electrostatic field overlap with the active ITO in the device region arising from bias. (b) Detail of the field line in plane with the ITO layer. The line plot represents the carrier concentration (blue-low and red-high)

etching of the oxide (Fig. 4). Here we use the aforementioned carrier concentration of $2x10_{20}$ cm-3 for ITO and $1x10_{14}$ cm-3 for Si, respectively. The plasmonic optical confinement in the active region further acts to amplify the material index change into obtainable effective modal index variation; hence aiding the overall modulation depth arising from both effects (traditional plasmonics and improved electrostatics in the lateral configuration). As both plasmonic top contact and the bottom contact in the capacitive stack are only metal paths, there is little resistance leading up to the device region; so, such a device is only limited by the capacitance (not R-limited) in terms of attainable speed. Selective plasma treatment on the ITO contact region can avail lower contact resistances up to 2 orders of magnitude [9]. The switching speed of such modulators are essentially limited by the dynamics of majority carriers in the ITO and, optimally, speeds in GHz ranges should be feasible as demonstrated in other majority carrier-based devices [63-65]. Footprint efficient modulators can also help in photonic-electronic hybrid integration for network on chips [66], and we envision emerging EO material heterogeneously integrated in foundry-ready photonic circuits to be a key driver in a modulator roadmap [67,68].

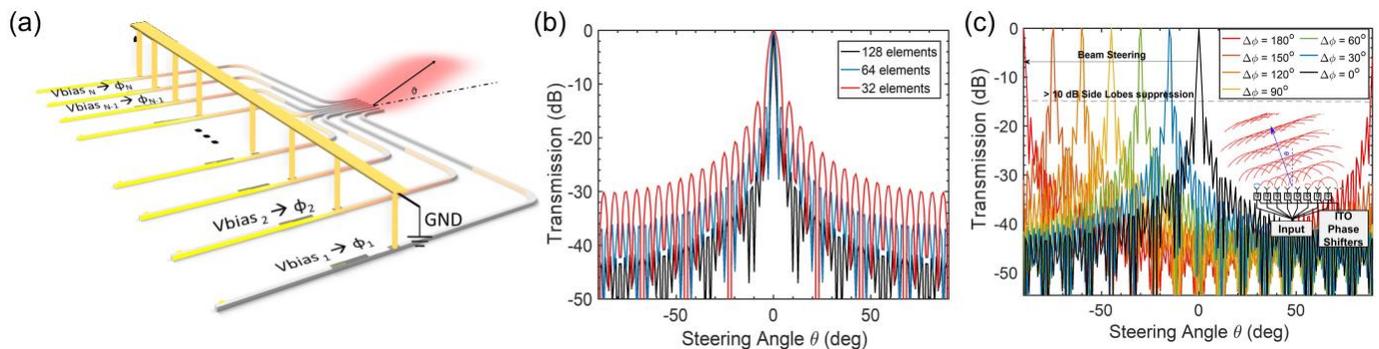

Fig. 5. (a) Schematic representation of an phased array end-fire beam steering platform using ITO lateral capacitor modulators; (b) Beam profile for a phased array which comprise of 32 (red), 64 (blue) and 128 (black) waveguides; and (c) Beam stearing radiation pattern for increasing phase variation between neighboring elements (red 180₀ to black 0₀).

## V. OPTICAL PHASED ARRAY

In this section, we proceed to numerically demonstrate a first design of an optical phased array system constituted by such ITO phase modulators using this hereby introduced lateral capacitor configurations (Fig. 5). The ITO-based phase modulators are used to finely control the phase of each channel. Ultimately, the waveguides approach each other in a 1-D array of end-fire antennas with sub-wavelength pitch ($\lambda/2$), hence emitting the beam from the edge of the chip. An overview of such a beam-steering scheme is illustrated in Fig. 5(a). With the phase tuning efficiency of the lateral-MOS capacitor ITO Mach Zehnder modulator characterized, it is possible to achieve phase shifts in plane with the waveguides (horizontal direction). The steering in $\psi$ for a uniform array is determined by $\sin \psi = \lambda \cdot \varphi/2\pi d$, where $\varphi$ is the uniform phase difference between the adjacent array elements, $\lambda$ is the free-space wavelength of the laser, and $d$ is the element spacing. Straightforwardly, the maximum steering angle can be achieved when $d$ equals to half the wavelength ($d = 775$ nm). From FTDT simulation and numerical modelling, the maximum steering angle achievable through electrostatic tuning of ITO based OPAs, for 8 waveguides, can reach $\pm 80°$ (Fig. 5c). Also, in this case, the maximum side lobe level (SLL) is more than 5 dB throughout the steering angle with an average beam width of 20°. However, this large beam width can be considerably decreased by augmenting the number of waveguides in the end-fire array. The SLL decreases significantly due to increased numbers of waveguides in the array and for 128 waveguides is larger than 10 dB Fig. 5(b). Moreover, in Fig. 5(c), the beam width (FWHM) decreases from ~20° to less than 3° by increasing the number of arrays from 8 to 128.

## VI. CONCLUSION

In conclusion, we have hereby demonstrated an MZI ITO-based electro-optic modulator, in a lateral capacitor configuration, which leads to favorable electrostatics and enhance the carrier tunability in the ITO film, thus delivering competitive performance in terms of figure of merit of $V_\pi L = 63$ V·µm for a <2 µm compact phase shifter. Using this modulator as a phase shifter as a building block of a 128-waveguides end-fire optical phased array beam steering platform, we find a rather fine narrow shaping of the main beam lobe (<3°) and >10 dB suppression of the side lobes, while steering up to ±80° emission profile. This approach has the potential to reduce loss in current OPA design, and yet availing of the ITO GHz-fast modulation speed, for the next-generation LiDAR system, holographic display, free-space optical communications, and optical switches.

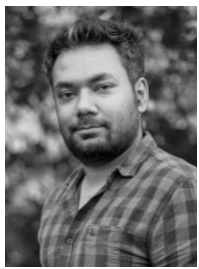

**Rubab Amin** received his B.S. degree with the distinction Summa Cum Laude in Electronics & Telecommunication Engineering from North South University, Dhaka, Bangladesh in 2011. Currently, he is a doctoral candidate and member of the OPEN Lab team at The George Washington University, Washington DC. His research is focused on design and demonstration of novel nanoscale electro-optic integrated modulators on Si-photonic platform. His research interests include nanophotonics, plasmonics, electro-optic modulators, transparent conducting oxides and low-dimensional material photonics.

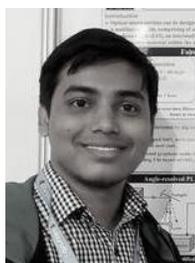

**Rishi Maiti** (M'19) received his M.Sc. degree in Physics from IIT Kharagpur, India in 2012, and then received his Ph.D. degree in 2017. His PhD research topic was studies on hybrid graphene nanostructures for optoelectronic devices. He joined the University of Brescia, Italy as a visiting scholar under Erasmus Mundus scholarship in 2016. Currently, he is a Post-Doctoral Fellow in the George Washington University. His research interests include the nanophotonic devices, Electro-optic modulator, optical interconnect, novel materials, plasmonics & metamaterials, Tunnel junction & Smart window, as well as their promising applications towards fully integrated photonic circuit.

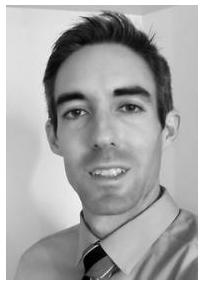

**Jonathan K. George** received the BS degree in computer science and applied mathematics from the University of Colorado at Colorado Springs in 2005, the MS degree in electrical engineering in 2015 from The George Washington University where he is currently pursuing the PhD in electrical engineering. He is a Staff Engineer at Northrop Grumman. His research interests include artificial intelligence, optical computing, and nonlinear optics.

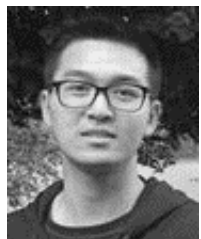

**Xiaoxuan Ma** (M'19) received the B.S. degree in Opto-Electronics Information Science and Engineering from Nanjing Tech University, Nanjing, China, in 2017, and the M.S. degree in electrical engineering from the George Washington University, Washington, DC, USA, in 2019. He is currently working toward the Ph.D degree from the Department of Electrical Engineering, George Washington University. His current research interests include electro-optic modulators, optical integrated circuit and LiDAR systems.

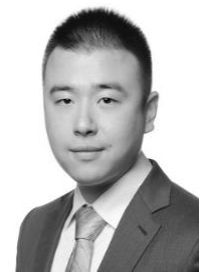

**Zhizhen Ma** (M'15) was born in Beijing, China, in1991. He received the B.S. degree in electrical engineering from the Huazhong University of Scienceand Technology, Wuhan, China, in 2013, and theM.S. degree in electrical engineering from the GeorgeWashington University, Washington, DC, USA, in 2015. He is currently working toward the Ph.D. de-gree from the Department of Electrical Engineering, George Washington University. His research interestsinclude nano-photonic devices, electro-optic modu-lators, 2D materials and metamaterial.

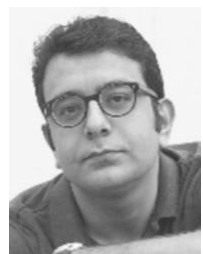

**Hamed Dalir** received his M.S. and Ph.D. degrees in electrical engineering from the Tokyo Institute of Technology in 2011 and 2014, respectively. Prior joining to Omega Optics, Dr. Dalir was a postdoctoral researcher at the Japan Society for the Promotion of Science (JSPS) and research associate at the University of California Berkeley in 2015 and 2016, respectively. His current research interests include design, fabrication, and characterization of III/V hybrid integrated nanophotonic silicon devices used in optical interconnects and RF sensing, including highly linear broadband graphene optical modulators, high-speed subvolt low-dispersion E.A modulator based on Bragg reflector waveguide, and transverse-coupled-cavity VCSELs. As the first author Dr. Dalir has published more than 100 peer-reviewed papers in prestigious journals and conferences. Dr. Dalir holds 6 issued and 4 pending patents on ultra-high speed devices as well as LiDAR. Dr. Dalir has served as a principle investigator for multiple projects supported by NASA, DoD and Broadcom. Dr. Dalir is a senior member of Optical Society of America (OSA),a senior member of IEEE and committee and session chair of SPIE.

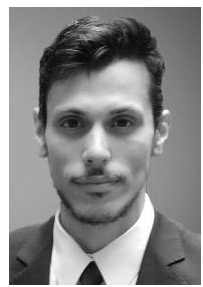

**Mario Miscuglio** is a post-doctoral researcher in Prof. Sorger's group in GWU. He received his Masters' in Electric and Computer engineering from Polytechnic of Turin, working as researcher at Harvard/MIT. He completed his PhD in Optoelectronics from University of Genova (IIT), working as research fellow at the Molecular Foundry in LBNL. His interests extend across science and engineering, including nano-optics and light-matter interactions, metasurfaces, Fourier optics and photonic neuromorphic computing.

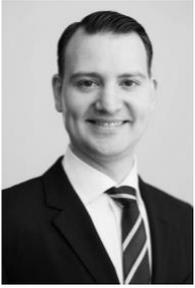

**Volker J. Sorger** (M'07) is an Associate Professor in the Department of Electrical and Computer Engineering and the leader of the Orthogonal Physics Enabled Nanophotonics (OPEN) lab at the George Washington University. He received his PhD from the University of California Berkeley. His research areas include opto-electronic devices, plasmonics and nanophotonics and photonic analog information processing and neuromorphic computing. Amongst his breakthroughs are the first demonstration of a semiconductor plasmon laser, attojoule-efficient modulators, and PMAC/s-fast photonic neural networks and near real-time analog signal processors. For his work, Dr. Sorger received multiple awards among are the Presidential Early Career Award for Scientists and Engineers (PECASE), the AFOSR Young Investigator Award (YIP), the Hegarty Innovation Prize, and the National Academy of Sciences award of the year. Dr. Sorger is the editor-in-chief of Nanophotonics, the OSA Division Chair for 'Photonics and Opto-electronics' and serves at the board-of-meetings at OSA & SPIE, and the scholarship committee. He is a senior member of IEEE, OSA & SPIE.